\documentclass{elsart}

\begin{document}
\begin{frontmatter}
\title{Leptons might not generate gravity}

\author{Homer G. Ellis\thanksref{email}}
\thanks[email]{{\it Email address:\/} Homer.Ellis@Colorado.EDU}
\address{Department of Mathematics, University of Colorado at Boulder,
  395 UCB, Boulder, Colorado 80309, USA}

\begin{abstract}
A simple thought experiment suggests that, contrary to assertions in an
earlier Letter, constancy across materials of the ratio of active to
passive gravitational mass does not rule out that electrons (and other
leptons) could have active gravitational mass zero, thus might not
generate gravity.  If they do not, then widely held assumptions about
the gravitational effects of various forms of energy cannot be
sustained.
\end{abstract}

\end{frontmatter}

A September 2001 Letter~\cite{Unni01} argues that improvements in the
sensitivities of certain experiments, that of Kreuzer~\cite{Kre68} in
particular, could settle the question whether leptons generate gravity.
The argument has two parts.  The first says that such improvements could
establish more firmly that the ratio of active gravitational mass to
passive gravitational mass (thus to inertial mass) is the same for all
material bodies.  (The Letter refers to the uniformity of this ratio as
`equality' of the masses, and takes the ratio to be~1, as can be
arranged by a suitable choice of unit for the active mass.)  The second
part of the argument, which is not made explicit, says that constancy of
this ratio across materials would require that not only the baryons but
also the leptons in atoms generate gravity.  The first part is
unexceptionable, but the second involves an unjustified hidden
assumption, which can be exposed in the following way.

As Eq.~(2) of the Letter is presented, it states that if the electrons
in a material body do not generate gravity, but everything else in it
does, then the active mass $M_a$ of the body is its passive--inertial
mass $M_{p,i}$ reduced by the sum $M^e_{p,i}$ of the passive--inertial
masses of its electrons: $M_a = M_{p,i} - M^e_{p,i}$.  Because the
experiments in question can at best confirm equality only between the
active mass and the passive--inertial mass of the \emph{whole body\/},
this equation is not justified --- but it is not the unjustified
assumption referred to above.  Let us replace Eq.~(2) by an equation
that would be justified by such confirmation of equality, namely that
$M_{p,i} = M_a = M^e_a + M^r_a$, where $M^e_a$ is the sum of the active
masses of the electrons and $M^r_a$ is the sum of the active masses of
the remaining constituents of the body (which masses are presumed to be
additive quantities).\footnote{\baselineskip 10pt The `additivity'
of active masses mentioned here, though precise in a linear theory of
gravity such as Newton's, must be understood in a nonlinear theory such
as Einstein's as referring to a nonlinear superposition of gravitational
effects that, at points an experimentally reasonable distance from the
body, can be treated as linear.  This is an assumption not treated in
the previous Letter and not to be treated in the present Letter; it too
is not the unjustified hidden assumption in question.}
Now the unstated part of the argument of the Letter, if made explicit,
would be that two homogeneous, geometrically congruent, electrically
neutral bodies B and $\mathrm B'$, made of different materials but
having by design $M_{p,i} = M_{p,i}'$, and by observation
$M_a = M_a'$, thus $M^e_a + M^r_a = {M^e_a}' + {M^r_a}'$, would have
nucleons in equal number, but different numbers of protons, thus
different numbers of electrons, thereby ruling out the possibility that
$M^e_a = {M^e_a}' = 0$, \emph{unless} the atomic and molecular binding
energies in B and $\mathrm B'$ generate gravity in just the amounts
required to balance the equation $M^r_a = {M^r_a}'$.  This is where the
unjustified assumption is hiding.

Consider, for clarity, the following thought experiment: A single,
isolated hydrogen atom, comprising one proton and one electron, is
approached by an antineutrino.  In a miraculous occurrence of reverse
$\beta$-decay the antineutrino grabs the electron and disappears with
it into the proton, thereby converting the hydrogen atom into a neutron.
If the electron and the antineutrino (leptons both) have active
gravitational mass zero, would the neutron's gravitational field differ
in any way from that of the hydrogen atom?  If so, then according to
conventional theory the difference must be attributed to changes in the
system's energy and passive--inertial mass.  But if electrons, which
have nonzero passive--inertial mass, don't gravitate, then the same may
be true of other manifestations of passive--inertial mass and, in light
of $E = m_{p,i} c^2$, of some forms of energy as well, in particular of
those that changed in the transition from hydrogen atom to neutron.  It
is thus perfectly consistent with the proposition that leptons do not
generate gravity to \emph{not\/} assume that the gravitational field
will differ for the hydrogen atom and the neutron.  That it will
differ is the hidden assumption in its barest form.

Let us extend this analysis to the Kreuzer experiment.  That experiment
compared the gravitational attraction exerted on test objects by each of
two homogeneous, geometrically congruent, electrically neutral bodies A
and B, differently constituted but weighing the same, thus having the
same passive--inertial mass $M_{p,i}$.  The precision of the
measurements allowed the inference that the ratios of active to
passive--inertial mass for the two bodies differed by less than
$5 \times 10^{-5}$.  Again for clarity, consider an idealized version of
the experiment in which body A is made of a single isotope of one
element, each of whose atoms has $p_{\mathrm A}$ protons, the same
number of shell electrons, and $n_{\mathrm A}$ neutrons, and body B is
made of a single isotope of another element, each atom of which has
$p_{\mathrm B}$ protons and shell electrons, and $n_{\mathrm B}$
neutrons, with
$p_{\mathrm A\!} + n_{\mathrm A} = p_{\mathrm B} + n_{\mathrm B}$
and $p_{\mathrm A} > p_{\mathrm B}$.  In each atom of body A, working
from the outermost electron shell inward, perform reverse beta decay by
stuffing \,$p_{\mathrm A} - p_{\mathrm B}$\, of its electrons, along
with as many antineutrinos, into its nuclear protons, thus turning
the protons into neutrons and the A atoms into B atoms, maintaining
congruence all the while.  Now the bodies are identically constituted
and their weights, therefore their passive--inertial masses, are still
the same.  But if neither leptons nor binding energies generate
gravity,\footnote{\baselineskip 10pt In the case of binding energies the
`ungenerated gravity' in question is the gravity outside the bodies,
where the electromagnetic field vanishes.  Within the bodies the
electromagnetic field can be nonzero, thus might generate internal
gravity not detectable externally.  This interpretation will be
maintained throughout the present Letter.}
then the active mass of body A before the transformation is the same as
that after the transformation, thus the same as that of B, and therefore
the ratio $M_a/M_{p,i}$ is the same for A~and~B --- despite that the
passive--inertial masses and the binding energies of A's atoms and
molecules have changed.  It is therefore the case that a
perfect-precision null result of the idealized Kreuzer experiment, and
by straightforward extension the actual experiment, cannot rule out that
leptons (and, concomitantly, binding energies) do not generate
gravity.

It is conceivable that, by themselves, the changes in the
passive--inertial masses and the binding energies of A's atoms and
molecules would have increased A's active gravitational mass, but that
this increase was exactly matched by a decrease owed to a change of
molecular kinetic energy necessary to maintain A's size, shape, and
weight.  It is also conceivable that they would have \emph{de}creased
A's active mass, and that this \emph{de}crease was compensated by a
change of kinetic energy.  It is, however, equally conceivable (and
from a probabilistic standpoint even more likely) that none of these
changes would cause any change in A's active gravitational mass.
Consequently, just as a null result of the Kreuzer experiment cannot
rule out nongravitating leptons, neither can it exclude that binding
energy and kinetic energy do not produce gravity.

A formulation of this conclusion that makes no reference to
transmutation of elements reads as follows: If
\vspace{-10pt}
\begin{itemize}
  \item[a.] two homogeneous, geometrically congruent,
            electrically neutral, material bodies of
            equal densities have the same total number
            of protons and neutrons, and
  \item[b.] every proton and every neutron, standing
            alone, would exhibit the same active
            gravitational mass as every other proton
            and every other neutron, and
  \item[c.] no constituent, material or otherwise, of
            either body other than its protons and
            neutrons generates any gravitational effect
            at a point an experimentally reasonable
            distance from that body, and
  \item[d.] whatever nonlinearities exist in the
            superposition of the gravity of the protons
            and neutrons of either body approximate
            those of the other body no less closely than
            do the nonlinearities in the superpositions
            of the electromagnetic fields generated by
            the bodies' constituents,
\end{itemize}
\vspace{-10pt}
then probing of the gravitational field at an experimentally reasonable
distance from either of those bodies would yield no information that
would allow one to decide which of the bodies was generating that field.

It is not simply that the Kreuzer experiment cannot rule out that
leptons, binding energy, and kinetic energy do not gravitate.  Rather it
is that such nongravitating is fully consistent with absolute, precise
constancy of the ratio of active to passive--inertial gravitational mass
across all material bodies composed of atoms and molecules with protons
and shell electrons in equal numbers, thus electrically neutral.  For
this reason the other experiments cited in~\cite{Unni01} as capable,
with improvements in precision, of demonstrating that leptons gravitate
cannot do so, as the most they can do is increase confidence in the
constancy of that ratio.  What an experiment using material bodies
carrying excess electric charge might show is, of course, a different
matter.

The notion that energy in all its forms produces gravity traces all the
way back to Einstein's 1916 paper Die Grundlage der allgemeinen
Relativit\"atstheorie~\cite{Ein16}. In that paper's~\S 16, titled in
translation The General Form of the Field Equations of Gravitation,
Einstein seeks a tensorial equation to correspond to the Poisson
equation $\nabla^2 \phi = 4 \pi \kappa \rho$, where $\rho$ denotes the
``density of matter''$\!$.  Drawing on the special theory of
relativity's identification of ``inert mass'' with ``energy, which finds
its complete mathematical expression in . . . the energy-tensor''$\!$,
he concludes that ``we must introduce a corresponding energy-tensor of
matter $\textstyle{T}^\alpha_\sigma$\,''$\!$.  Further describing this
energy-tensor as ``corresponding to the density $\rho$ in Poisson's
equation''$\!$, he goes on to invent the field equation that bears his
name:
$R_{\mu \nu} - \half R \, g_{\mu \nu} = -8 \pi \kappa T_{\mu \nu}$, as
currently expressed.  Here Einstein confounded `gravitat{\em ing\/}
mass', which is the sole contributor to the ``density of matter'' in
Poisson's equation, with ``inert mass''$\!$, thus with energy by way of
$E = m c^2$ and with `gravitat{\em ed\/}' mass by way of the equivalence
between inertial mass and passive gravitational mass.  Whether such a
confounding can be justified by experimental evidence is the underlying
question addressed by the previous Letter~\cite{Unni01} and this
one.\footnote{\baselineskip 10pt That Einstein confounded active mass
with passive--inertial mass, knowingly or unknowingly, is borne out
further by the statement in his~\S 16 that for a ``complete system
(e.g.~the solar system), the total mass of the system, and therefore its
total gravitat{\em ing} action as well, will depend on the total energy
of the system, and therefore on the {\em ponderable} energy together
with the gravitational energy.'' (Emphases added.)  Let it be
remembered, however, that he thought of the field equation with the
``energy-tensor of matter'' in it as similar to a building with one wing
made of fine marble (geometry) and the other of low-grade wood
(energy-tensor), which ultimately should be replaced by an equation
whose architectural analog would consist of marble alone~\cite{Ein50}.}

If confounding of active gravitational mass with passive--inertial mass
and with energy is not justified, then Einstein's field equation is open
to modification in two ways: a) some of the forms of energy usually
included in the tensor $T_{\mu \nu}$ can be left out; b) the couplings
to geometry of those forms of energy left in can differ from the usual.
In a paper that appeared some thirty years ago~\cite{Ell73}, exercise of
option (b) to reverse the polarity of the coupling of a scalar field to
geometry produced a space-time manifold which was described
in~\cite{Ell73} as a `drainhole', and which has since been recognized as
an early, perhaps the first example of what is now called a `traversable
wormhole'~\cite{Mor88,Cle89}.  Specifically, with $\phi$ governed by the
wave equation $\Box \phi := \phi^{,\kappa}\!{}_{;\kappa} = 0$
and scaled so that
$T_{\mu \nu} = (1/4 \pi \kappa)(\phi_{,\mu} \phi_{,\nu} -
\half \phi^{,\kappa} \phi_{,\kappa} \, g_{\mu \nu})$,
Einstein's equation
%$R_{\mu \nu} - \half R \, g_{\mu \nu} = -8 \pi \kappa T_{\mu \nu}$
was replaced by
$R_{\mu \nu} - \half R \, g_{\mu \nu} = 8 \pi \kappa T_{\mu \nu}$.
Conventional wisdom
says that this coupling somehow makes the energy of the scalar field be
\emph{negative}, and that the scalar field must therefore be associated
with `exotic' matter.  But if, as argued here, the coupling of energy to
geometry is not dictated by experimental observation, then one can just
as well say that the energy of the scalar field is positive, that the
reversed-polarity coupling is as justifiable as the conventional
coupling, and that nothing `exotic' is involved.  This is clearly
apparent in~\cite{Ell73}, where the gravitational field is untangled
from the geometry of space, and the scalar field is seen to be coupled
essentially to the geometry of space alone, even to the extent that
gravity can be turned off completely while the drainhole stays open.
The effect of the reversed polarity of the coupling is to let in the
negative spatial curvatures that must be present if a stable traversable
wormhole throat is to exist.  With or without gravity turned on, the
space-time is horizonless, geodesically complete, and singularity-free,
the Penrose--Hawking singularity theorems~\cite{Haw73} having been
escaped by denial of their primary hypotheses.

A further demonstration of the reasonableness of the reversed-polarity
coupling of the scalar field to geometry occurs in~\cite{Ell79}, which
extracts from the field equations a metric describing a nonstatic,
nongravitating drainhole--traversable-wormhole whose throat, starting
with infinite radius in the infinitely distant past, chokes down to a
single point, instantly reopens, then expands back to infinite size in
the infinitely distant future.

Apropos of option (b), a relatively recent paper~\cite{Lo97} has argued
that not all types of energy are equivalent to mass, and that those that
are not, such as electromagnetic energy, can couple to geometry in ways
different from the way that mass couples to it.

As to option (a), if the presumption that kinetic energy generates
gravity is not justified, then the same should be true for pressure in
a fluid or a gas.  This allows the usual mass-energy-stress-momentum
tensor to be replaced by one without pressure terms.  That produces a
solution analogous to but simpler than the Schwarzschild interior
solution, which I shall describe in a subsequent paper.

Lastly I would for historical purposes point out that, as noted in the
previous Letter, the question whether electrons generate gravity
was first posed explicitly twenty years ago~\cite{Jac83}, if not
earlier.

\end{document}